\documentclass{ws-ijmpe}

\begin{document}

\markboth{G.B. Adera, B.I.S. Van der Ventel}{Associated Hyperon-Kaon Production via Neutrino-Nucleus Scattering}

\catchline{}{}{}{}{}
\title{ASSOCIATED HYPERON-KAON PRODUCTION VIA NEUTRINO-NUCLEUS SCATTERING}

\author{\footnotesize G.B. ADERA}

\address{Department of Physics, University  of Stellenbosch, Matieland,\\
Stellenbosch 7602, South Africa \\
adera@sun.ac.za}

\author{\footnotesize B.I.S. VAN DER VENTEL}

\address{Department of Physics, University  of Stellenbosch, Matieland,\\
Stellenbosch 7602, South Africa \\
bventel@sun.ac.za}

\maketitle

\begin{history}
\received{(received date)}
\revised{(revised date)}
\end{history}

\begin{abstract}
We present the investigation of the neutrino-induced strangeness associated production on nuclei in the relativistic plane wave impulse approximation (RPWIA) framework at the intermediate neutrino energies. In this study, the elementary hadronic weak amplitudes are embedded inside the nuclear medium for the description of the exclusive channels of neutrino-nucleus interactions. These amplitudes are extracted using a model-dependent evaluation of the hadronic vertex using the Born term approximation in which the application of the Cabibbo V-A theory and SU(3) symmetry are assumed to be valid. The nuclear effects are included via the bound state wavefunctions of the nucleon obtained from the relativistic mean field (RMF) models. Two kinematics settings are used to examine various distributions of the differential cross section in the rest frame of the target nuclei. The numerical results are obtained for the neutrino-induced charged-current (CC) \,$\rm K^{^+}\Lambda$-production on bound neutrons in $1s^{1/2}$ and $1p^{3/2}$ orbitals of  $^{12}$C. The angular distributions are forward peaked under both kinematic settings, whereas under the quasifree setting the cross sections tend mimic the missing momentum distribution of the bound nucleon inside the nucleus.
\end{abstract}

\section{Introduction}\label{sec01}

The study of neutrino interactions, in general, has played a tremendous role in testing the validity of various models of the electroweak unification theory. In particular,  the weak neutrino-nucleus interactions involving exclusive associated production of strange particles in the intermediated energy region have become one of the major theoretical and experimental topics in the fields of astrophysics, cosmology, particle and nuclear physics. At the fundamental level of the standard model the theoretical and experimental investigations of these exclusive channels assist in gaining new and better insight into nucleon and nuclear structures. As such, neutrino interactions can play a pivotal role by complementing the theories of particle and nuclear physics that have been mainly established based on the electromagnetic interactions. Above all, the study of neutrino-induced strangeness associate production provides an improved understanding of basic symmetries of the standard model, strange-quark content of the nucleon, structure of the weak hadronic form factors, strong coupling constants, and medium modification of the elementary amplitudes. Therefore, strange particle production via the weak neutrino-nucleus interaction is worthy of further theoretical investigation on its own, even more than that due to the role it plays as a platform where the studies of particle and nuclear physics can overlap.

Precision neutrino physics necessitates a detailed theoretical knowledge of neutrino-induced associated production of strange particles on nuclei owing to the presence of such channels as backgrounds in the detection of the neutrino-oscillation signal. Along this line MINER$\nu$A\cite{MINERVA}, K2K\cite{K2K}, and MiniBooNE\cite{MiniBOONE} experiments are reportedly gathering high-statistic data on neutrino-oscillation signals as well as exclusive kaon-hyperon production cross sections. There are also other relevant backgrounds like the quasi-elastic (QE) and pion production channels which have been well studied because of their relative simplicity and comparatively bigger cross sections. Therefore, these backgrounds need to be carefully addressed in the analysis of the neutrino-oscillation data at low and intermediate neutrino energy regime.\cite{Minerva-prop} In those experiments the nuclei that have been used as targets are $^{12}$C, $^{16}$O, and $^{208}$Pb, just to mention a few. On the other hand, beyond the standard model the supersymmetric grand unification theories (SUSY GUTs) clearly emphasizes the significance of a thorough understanding of neutrino-induced strangeness associated production backgrounds pertaining to the fact that their subsequent decay processes, when initiated by atmospheric neutrino fluxes, mimic the nucleon decay signal.\cite{WAM03,NiSo05} Currently, there is an ongoing proton decay search being carried out by using the world's largest nucleon decay detector Super-KamioKande,\cite{Super-K} and future proposed plans such as LAGUNA\cite{LAGUNA} and also other next generation experiments involving underground Megaton detectors such as Hyper-KamioKande or UNO.\cite{Hyper-K} The success of these experiments depends on how they will be able to separate the backgrounds from the decay signal despite the effort to minimize their impact.

Early experimental observations of the neutrino-induced exclusive final states, in which a kaon belonging to a  pseudoscalar meson octet is produced in conjunction with a hyperon in the baryon octet, were reported at ANL\cite{ANL}, CERN\cite{CERN1,CERN2,CERN3}, and BNL\cite{BNL}. The data from those experiments, however, are known for their low-statistics and large systematic error. Moreover, previous rough theoretical estimations done by Shrock\cite{PRC.D12.2049.1974}, Mecklenburg\cite{APA48.293.1978}, Amer\cite{PRC.D18.2290.1978}, and Dewan\cite{PRC.D24.2369.1981} were based on the assumption that nucleon targets are free particles. In actual experiments, however, the nucleons are bound inside the target, hence invoking the necessity of taking into account the medium modification effects such as Fermi motion, Pauli blocking, and binding energy correction. As a consequence, a proper theoretical and experimental comparison has never been done owing to those uncertainties incurred on both sides.\cite{PRC.D24.2369.1981} 

In the intermediate energy region, neutrino-nucleus interactions involves  exclusive channels of either one or two strange particles. The final states of semileptonic strangeness production can be categorized in terms of  the change in charge and the change in strangeness quantum numbers of the weak hadronic transition current.  The first category comprises the CC and strangeness conserving ($\Delta S = 0$) processes in which either a K$^0$ or K$^+$ (S =1) meson is produced in conjunction with either a $\Lambda$, $\Sigma^0$, or $\Sigma^{\pm}$ (S = -1) hyperon. The bubble chamber experiments detected events of the $\nu_{\mu}\rm{n} \rightarrow \mu^- K^+ \Lambda$ channel from this category. The second category is distinguished by CC and $\Delta S = 1$ currents and here the $\Delta S = \Delta Q$ selection rule is applicable, as a consequence only single strange particle $K^+$ or $K^0$ is produced in the final state. Some of the observed channels under this category are $\nu_{\mu}\rm{p} \rightarrow \mu^- K^+ p$ and $\nu_{\mu}\rm{n} \rightarrow \mu^- K^0 p$. Neutrino exclusive processes involving the neutral-current (NC) and $\Delta S = 0$ hadronic weak transition form the third category and the observed channels in the bubble chamber experiment are $\nu_{\mu}\rm{p} \rightarrow \nu_{\mu}K^+\Lambda$ and $\nu_{\mu}\rm{n} \rightarrow \nu_{\mu} K^0 \Lambda$. So far there has not been any observation of exclusive channels of the forth category (i.e., NC and $\Delta S = 1$  processes) and henceforth their detection in the ongoing and future neutrino experiments will have a great implication for the emergence of new physics beyond the standard model.\cite{WAM03,NiSo05} 

The detailed data with minimum statistical error for heavier nuclei will be available from MINER$\nu$A and other ongoing and planned experiments. Moreover, the renewed effort of theoretical formulation of the neutrino-induced associated production that incorporates the nuclear effects will allow the comprehensive analysis of experimental data using models. Motivated by the ever growing experimental interests, in our recent work\cite{PRC82.025501.2010}  we have revived a few long-forgotten theoretical studies of exclusive strangeness associated production via $\nu-\rm{N}$ interaction with particular interest in CC and $\Delta S =0$ channels. The paper introduced a new technique of extracting weak hadronic elementary amplitudes from a model-dependent evaluation of the hadronic vertex in the framework of electroweak theory and fundamental symmetries of the standard model. The most general nature of our formulation makes it very convenient to be implemented in the studies of the CC and $\Delta S = 1$ and NC and $\Delta S =0$ channels as well quite easily,  without having to make significant alterations. The model of the current study is constructed on the basis of our previous work done at elementary particle level, and aimed at accounting for nuclear medium effects. On the other hand, Rafi Alam $et\, al$\cite{PRC82.033001.2010} have investigated the CC and $\Delta S = 1$ processes, in which a single strange particle ($S = 1$) is produced, by employing the effective Lagrangian formulation in the chiral symmetry framework. Therefore, a meaningful interpretation of data and an improved understanding of hadronic and nuclear weak physics with particular interest in the Cabibbo V-A theory, the SU(3) symmetry, and medium modification effects will be possible in the near future. 

It is important to notice the fact that neutrino-nucleus interactions at neutrino energy in the order of few GeV also open up to other final state channels, for instance, a single strange and non-strange meson productions which may even have higher probability of detection. Nevertheless, our effort in the theoretical investigation of the neutrino-induced strangeness associated production of pseudoscalar meson and hyperon is motivated and also strongly supported by early experimental observations and the present and future experimental activities that we mentioned earlier. On the contrary, near threshold, the neutrino-induced associated production of vector kaon with identical quark structure as pseudoscalar kaon has never been reported, to our knowledge. This may partly be due to their large masses and relative instability. Thus the invariant mass of the final state of strange vector meson and hyperon requires a minimum neutrino energy that may exceed the near threshold energy limit upon which our study and those accelerator and non-accelerator based experiments are focused. In addition, their instability, leading to very short lifetime, makes their experimental observation more difficult than those lighter and somewhat stable pseudoscalar kaons. Moreover, at low and intermediate neutrino energies nuclear physics studies have greater interest in pseudoscalar mesons than vector mesons. From the theoretical perspective, the conventional derivation of the invariant matrix element of the associated production has been simplified under the pseudoscalar meson field treatment. But for the channels, if any, having the vector meson and hyperon fields in the final state, the formulation presents a bigger complications and is beyond the scope of this study. 

In this paper a fully relativistic formalism for the production of strange particles from nuclei via the weak interaction is presented. The main aim of this formalism is to offer a relativistic description in terms of the angular, missing momentum, and  energy distributions of the differential cross section in the few GeV neutrino energy region. We develop a model, that makes use of elementary weak amplitudes, by invoking the impulse approximation at the hadronic interaction vertex. We assume that the incident neutrino interacts, via the exchange of a single vector boson, with a bound nucleon inside the target nucleus and then scatters into a final state lepton. On the hadronic part, the electroweak interaction of the weak gauge boson with the bound nucleon opens a final state channel involving the production of a pseudoscalar meson in association with a hyperon. In this approximation the rest of the nucleons behave like spectators and make up the residual nucleus. In this specific calculation all external particles will be regarded as free (on-shell) particles and subsequently described by plane wavefunctions except for a  bound (off-shell) nucleon. The motivation behind this will be expanded upon later. 

In our calculations, the RMF approximation to the Walecka model with the NL3 parametrization is employed due to the fact that, unlike a relativistic Fermi gas (RFG) model, the nuclear effects such as Fermi motion, Pauli blocking, and nuclear binding are naturally incorporated via the nucleon bound state wavefunctions. On the other hand, the elementary hadronic weak amplitudes are extracted from the Born term approximation, in which we assume that the standard Cabibbo V-A theory along with the conserved vector current (CVC) hypothesis as well as the SU(3) symmetry estimates of the strong coupling constants are valid and applicable. In the dynamical description of the weak hadronic vertex, we only consider the lowest-order Feynman diagrams of $s$-, $u$- and $t$-channels mediated by hadrons belonging to a baryon octet and a pseudoscalar octet. This approximation is valid near threshold where strong resonances are assumed to be absent except for the lowest-lying ones, which in fact decay mainly into N$\pi$ and N$\pi\pi$.\cite{PRC.D12.2049.1974}  The numerical results of our formalism for a typical neutrino-induced CC reaction process, in which $\rm K^{^+}$ is produced in conjunction with  $\Lambda$ on a neutron bound in the $^{12}$C nucleus, are presented under different kinematic settings.   

This paper is organized as follows. In section \ref{sec02} we present the relativistic description for the neutrino-induced CC associated production on nuclei, whereby we incorporated the nuclear structure effects through the bound state wavefunction. In section \ref{sec03} we develop our model for the nuclear weak transition operator in terms of the elementary weak current operator by invoking impulse approximation, and then evaluate the RPWIA cross section. In section \ref{sec04} we provide a model-independent parametrization of the hadronic current operator in terms of a set of invariant amplitudes and also explain main sources of uncertainty in our formulation as well. In order to obtain numerical results we calculate these invariant amplitudes by employing the Born term approximation. In section \ref{sec05} we present the numerical results for the RPWIA calculations under different kinematic settings. Finally, in section \ref{sec06} we summarize our results and provide topics for future investigation. 

\section{Formalism}\label{sec02}

\begin{figure}[\h]
\centering
\includegraphics[width=7.25cm, height= 5.75cm]{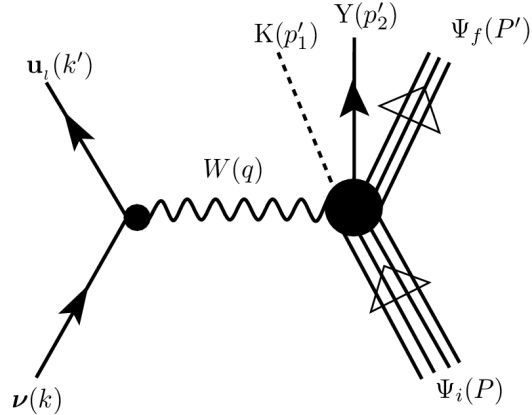}  \vspace*{4pt}
\caption{The lowest order Feynman diagram for the neutrino-induced associated production of strange particles on nuclei. The incoming neutrino interacts with the target nucleus via the exchange of a gauge boson, and subsequently the final state lepton, kaon, and hyperon are produced and then exit the reaction vicinity, leaving behind the residual nucleus.}
\label{fig01} 
\end{figure}

In this section we present the formalism for the relativistic neutrino-nucleus scatting processes. The reaction processes under consideration are of the form
\begin{equation} \label{Eqf-01}
 \nu(k) + \mathrm A(P) \rightarrow  l(k') + \mathrm K(p'_1) + \mathrm Y(p'_2) + \mathrm B(P'),
\end{equation}
as shown in Fig. \ref{fig01}. The incoming neutrino with four momentum $k^{\mu} = (E_k,\,\mathbf k )$ scatters into the final lepton $l$, with $k'^{\mu}=(E_{k'},\,\mathbf{k}')$ off the target nucleus $\rm A$, with  $P^{\mu} = (E,\,\mathbf{P})$ via the exchange of a single weak vector boson $W = W^{^\pm}$ or $Z^{^0}$, that carries the four-momentum transfer $q^{\mu} = (\omega,\,\mathbf{q})$. 
In the exclusive strangeness associated production channel, the final state involves the production of a pseudoscalar meson $\rm K$, with ${p'_1}^{\mu} = (E_{p'_1} ,\,\mathbf{p}'_1)$ accompanied by hyperon $\rm Y$, with ${p'_2}^{\mu} = (E_{p'_2},\,\mathbf{p}'_2)$ and then both exit the reaction vicinity, leaving behind a recoiling residual nucleus $\rm B$, whose four momentum is labeled as $P'^{\mu}=(E',\,\mathbf{P}')$. We also denote the masses of the target nucleus, final lepton, kaon, hyperon, and residual nucleus as $M$, $M_{_l}$, $M_{_K}$, $M_{_Y}$, and $M'$, respectively. 

\subsection{Differential cross section}

The theoretical description of the process is given in the laboratory frame in which the target nucleus is at rest, and hence its four-momentum becomes $P^{\mu} = (M,\,\mathbf{0})$. We follow the standard procedure of constructing the differential cross section.\cite{RQM-BjDr-1964} Thus, for the neutrino-nucleus reaction that proceeds through the $\rm KY$-production channels as shown in Fig. \ref{fig01} we may write down the most general differential cross section as
\begin{equation}\label{eqdxs01}
\begin{array}{ll} \mathrm d \sigma & = 
\dfrac{\mathcal{C}_1\mathcal{C}_2}{|\mathbf{v}_1 - \mathbf{v}_2|} (2\pi)^4 \delta ^4 (k + P - k' - p'_1 - p'_2 - P') \vert \mathcal M_{fi} \vert^2  \\
& \times \mathcal{C}_3 \dfrac{\mathrm d ^3 \mathbf{k}'}{(2\pi)^3}\mathcal{C}_4 \dfrac{\mathrm d ^3 \mathbf{p}'_1}{(2\pi)^3} \mathcal{C}_5 \dfrac{\mathrm d ^3 \mathbf{p}'_2}{(2\pi)^3}\mathcal{C}_6 \dfrac{\mathrm d ^3 \mathbf{P}'}{(2\pi)^3},
\end{array}  
\end{equation}
where $|\mathbf{v}_1 - \mathbf{v}_2|$ is the relative velocity of the incident neutrino with respect to the target nucleus, and in a rest frame of the target nucleus its value becomes unity owing to the fact that we neglect the (very small) neutrino mass. 
The overall four-momentum conservation is maintained by the four-dimensional $\delta$-function. $\mathcal{M}_{fi}$ is the invariant matrix element that contains the dynamical information of the weak nuclear interaction vertex. Moreover, the expression contains the kinematic phase space factor of the final states. Note that $\mathcal{C}_i$ are volume normalization factors that can be specified according to the Bjorken and Drell convention.\cite{RQM-BjDr-1964} 

\subsection{Relativistic impulse approximation}

\begin{figure}[th] 
\centering
\includegraphics[width=11.00cm, height= 5.50cm]{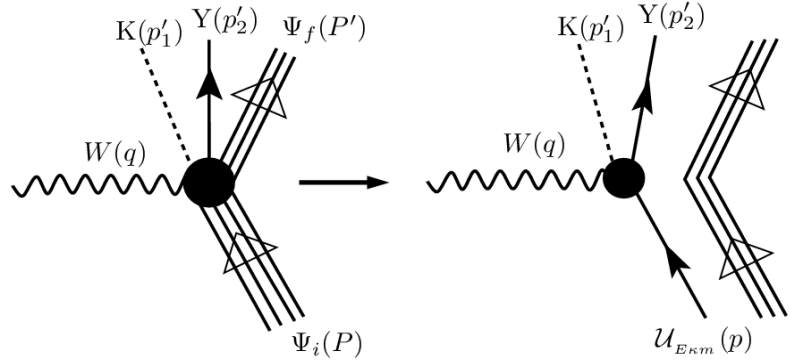} \vspace*{4pt}
\caption{The vertex approximation of the associated $\rm KY$-production vertex. Here the vector gauge boson interacts with a single bound nucleon inside the target nucleus, and subsequently the final state kaon and hyperon are produced, whereas the rest of the nucleons behave as spectators.}
\label{fig02}
\end{figure}

The main focus of this paper will be the derivation of the cross section for the associated production of strange particles in the framework of a relativistic impulse approximation (RIA). One of the RIA assumptions is that the production process takes place via the interaction of a single gauge boson with a bound nucleon inside the nucleus as shown in Fig. \ref{fig02}. In addition we make the impulse approximation which assumes that the
in-medium nuclear current operator can be replaced by the free nuclear current operator. 
It is therefore clear that the model consists of embedding the underlying free process in the nuclear medium. However, this immediately introduces the problem of nuclear medium effects which must be circumvented by making suitable and well-motivated assumptions. The bound state wavefunction of the nucleon are obtained from the modified Walecka model by invoking the RMF approximation.\cite{PRC55.540.1997} In general, there are two different frameworks that we can work with in the impulse approximation, and in this work we only consider the RPWIA that will be discussed in section \ref{sec03}. Note also, that such calculation can be generalized to the relativistic distorted wave impulse approximation (RDWIA) framework, which falls beyond the scope of this paper.

\subsection{Kinematics}

It is well established that in the relativistic description of the semileptonic weak interaction, the differential cross section can be separated into kinematic and dynamic parts. This, in turn, permits the treatment of these two parts individually without loss of generality. In what follows we briefly discuss the degrees of freedom we have at our disposal and the constraints that naturally emerge to minimize them.
\begin{figure}[\h]
\centering
\includegraphics[width=13cm,height=5.5cm]{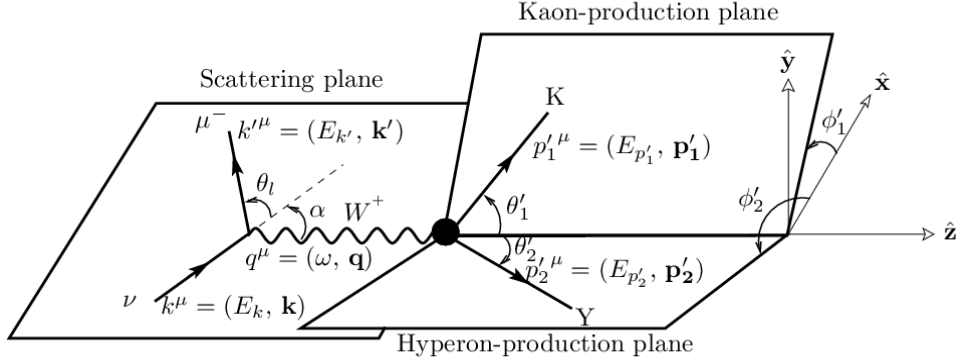} 
\caption{The kinematics of the charge current neutrino-induced associated production in the rest frame of the target nucleus. The geometry of the production process is depicted in terms of scattering, kaon-production, and hyperon-production planes. For convenience sake the recoil of the residual nucleus is left out of this geometric representation of the kinematics.}
\label{figkin01} 
\end{figure}
We first define the coordinate system as illustrated in Fig. \ref{figkin01}. The $z$-axis is set along the three-momentum transfer (i.e., $\hat{\bf{z}} =  \hat{\bf{q}}$). The lepton scattering plane can be used to specify the other two coordinate axes: $\hat{\bf{y}} = \hat{\,\bf{k}}\times \hat{\,\bf{k}}'/\sin \theta_l$, where $\theta_l$ is the lepton scattering angle, and consequently the $x$-axis is constrained to the leptonic plane and determined using $\hat{\bf{x}} = \hat{\bf{y}} \times \hat{\bf{z}}$. Recalling that we are working in the rest frame of the target nucleus our choice of coordinates gives rise to the vanishing $y$-components of any three-momenta constrained in the scattering plane. Moreover, at the leptonic transition vertex $\omega = E_k - E_{k'}$ and $\mathbf q = \mathbf k - \mathbf k'$. 

The leptonic weak transition presents the challenge of determining the four-momenta $k^{\mu}$, $k'^{\mu}$, and $q^{\mu}$ given that the kinematic variables $E_k$, $\omega$, and $|\mathbf q|$ are fixed. Thus one needs to impose energy-momentum conservation at the leptonic vertex and also invoke the on-shell conditions on the neutrino and final state lepton in order to determine the three-momenta:
\begin{eqnarray}\label{eqkin01}
k_x = |\mathbf k|\sin \alpha ,\,\,\,\,\,\,\,\,\,\,\,\,\,\,\,\,\,\,k_y = 0,\,\,\,\,\,\,\,\,\,\,\,\,\,\,\,\,\,\,\, k_z =  |\mathbf k|\cos \alpha \,\,\,\,\,\,\,\,\,\,\,\,\\
k'_x = |\mathbf k'|\sin (\alpha + \theta_l) ,\,\,\,\,\,\,\,\,k'_y = 0,\,\,\,\,\,\,\,\,\,\,\,\, k'_z =  |\mathbf k'|\cos (\alpha + \theta_l),
\end{eqnarray}
where 
\begin{equation}\label{eqkin02}
\sin \alpha  =   \dfrac{|\mathbf k'|}{|\mathbf q|} \sin \theta_l,\,\,\,\,\,\,\, \cos \alpha = \dfrac{1}{|\mathbf q|}[|\mathbf k| - |\mathbf k'|\cos \theta_l],
\end{equation}
and 
\begin{equation}\label{eqkin03}
\cos \theta_l \approx \dfrac{q^2 + 2E_k E_{k'} - M_{_l} }{2E_k|\mathbf k'|} ,\,\,\, E_{k'} \gg M_{_l}.
\end{equation}
Note that we make use of relativistic relations: $E_k = |\mathbf k|$, $E^2_{k'} = |\mathbf k'|^2 + M^2_{_l}$, and $q^2 = \omega^2 - \mathbf q^2$. In the Feynman diagrammatic formulation the weak gauge boson is identified as an internal line, as a consequence it shall remain off-shell (i.e., $q^2 \neq M^2_{_W} $). 

The nuclear vertex in Fig. \ref{fig01} represents the interaction of the weak gauge boson with the target nucleus and then resulting in the associated production of strange particles. In the final state, in addition to the kaon and hyperon, we also have the recoiling residual nucleus. This leads to the general energy-momentum conservation: 
\begin{equation}\label{eqkin04}
\omega + M = E_{p'_1} + E_{p'_2} + E',\,\,\,\,\,\,\,\,\,\mathbf P' = \mathbf q - \mathbf p'_1 - \mathbf p'_2,
\end{equation}
where $E' = M' + T_m$ and we assume that the recoil kinetic energy $T_m = \mathbf p_m^2/2M'$, of the residual nucleus is small enough to be neglected. Note that $\mathbf p_m = \mathbf P'$ is the missing momentum. The mass of the residual nucleus is obtained from $M' = M - (M_{_N} - E_b)$, where $M_{_N}$ is the nucleon mass and $E_b$ is the binding energy of the nucleon. Note that the binding energy can be determined from different nuclear structure models and in this particular work we use the values from the RMF model with the NL3 set of parameters.

Turning our attention to Fig. \ref{fig02} in which the vertex approximation is applied to the weak transition of the nucleus, we notice that the exchange gauge boson is absorbed by a single bound nucleon with four-momentum $p^{\mu}=(E_{a},\,\mathbf p_{_a})$, and subsequently the kaon and hyperon are produced in their respective planes specified by the azimuthal angles $\phi'_1$ and $\phi'_2$ as shown in Fig. \ref{figkin01}. In this reaction picture the rest of the nucleons behave as spectators. Furthermore, energy-momentum conservation leads to:
\begin{equation}\label{eqkin05}
E_{a}  = E_{p'_1} + E_{p'_2} - \omega,\,\,\,\,\,\,\,\,\,\mathbf p_{_a} = - \mathbf P'. 
\end{equation}
It is also worth noting that at the nuclear vertex, the target nucleus, residual nucleus, outgoing kaon and hyperon are on their respective mass shell except for the bound nucleon. This is what causes the off-shell ambiguity in connection to elementary process. In summary, the choice of a proper reference frame, specifying the coordinates, enforcing four-momentum conservation, and invoking the on-shell conditions are used to minimize the number of independent kinematic variables needed to describe the reaction process. One also needs to choose a proper orientation of the production planes with respect to the scattering plane in order to obtain reliable results. 

\subsection{Nuclear structure models}

In order to account for the nuclear structure effects, we resort to the RMF approximation to the Walecka model. In this model, the bound nucleon is described by the exact and self-consistent Dirac-Hartree solutions obtained from the effective Lagrangian constructed in the framework of the improved Walecka model. Walecka introduced the first quantum hydrodynamics model which is also known as QHD-I.\cite{ANP83.491.1974} Since then other models such as QHD-II,\cite{ANP16.1986,BDJD86} NL3,\cite{PRC55.540.1997}, and FSUGold\cite{PRC67.044317.2003͒} have added more complexities to the effective Lagrangian in the RMF approximation to improve the model's predictive power, for instance, in reproducing the ground-state properties of nuclei with more accuracy. In this work, therefore, we use the NL3 parameter set to generate the bound state wavefunction of the nucleon, owing to the fact that it has been known to be the most successful model in describing the ground-state nuclei reasonably well so far. 

\subsubsection{Bound state wavefunctions}

By invoking the RMF approximation to the Walecka model the position space bound state wavefunction can be obtained:  
\begin{equation}\label{eqrns01}
\mathcal{U}_{_{E\kappa m}}(\mathbf{x}) = \dfrac{1}{x} \left( \begin{array}{l}
g_{_{E\kappa}}(x)\mathcal{Y}_{\kappa m}(\hat{\mathbf{x}})\\
\\
if_{_{E\kappa}}(x)\mathcal{Y}_{-\kappa m}(\hat{\mathbf{x}})
\end{array}\right),
\end{equation}
where $E$ is the binding energy, $\kappa$ the generalized angular momentum, which uniquely specifies both the orbital angular monument $l$, and the total angular momentum $j$ where $j = |\kappa| - 1/2$; $l =  \kappa$ for $\kappa > 0$  or  $l = -1 - \kappa$ for $\kappa < 0$. $m$ is the projection of $j$.  $\mathcal{ Y}_{\kappa m}(\hat{\mathbf{x}})$ is the spin-angular wavefunction that couples the Pauli spinor $\boldsymbol \chi_{s_{z'}}$, with the spherical harmonics $Y_{l,m_{l}}(\hat{\mathbf{x}})$, of order $l$:
\begin{equation}\label{eqrns02}
\mathcal{Y}_{\kappa m}(\hat{\mathbf{x}}) = \sum_{s_{z'} = \pm \frac{1}{2}} \langle l,\frac{1}{2},m-s_{z'},s_{z'} \vert j\, m \rangle \, Y_{l, m - s_{z'}}(\hat{\mathbf{x}})\,\boldsymbol \chi_{s_{z'}}
\end{equation}
where $\langle l,\frac{1}{2},m-s_{z'},s_{z'} \vert j\, m \rangle$ are the Clebsch-Gordon Coefficients. 
Note that  $g_{_{E\kappa}}(x)$ and $f_{_{E\kappa}}(x)$ are not analytical functions but are generated numerically and then must be stored in large arrays. Moreover, the minus sign assigned to $\kappa$ in the lower component of the bound state wavefunction has to do with reproducing the exact value of the corresponding orbital angular momentum $l' = 2j-l$. 

In the plane wave case (see later) one needs the momentum space bound state wave function
which is the Fourier transform of the wavefunction in Eq. (\ref{eqrns01}) and is given by    
\begin{equation}\label{eqrns03}
\mathcal{U}_{_{E\kappa m}}(\mathbf{p}) = \int \, \mathrm{d^3}\mathbf{x}\, e^{-i \mathbf{p}\cdot \mathbf{x} }\,\mathcal{U}_{_{E\kappa m}}(\mathbf{x}).
\end{equation}

To determine an explicit expression for $\mathcal{U}_{_{E\kappa m}}(\mathbf{p})$ we employ the partial wave
expansion
\begin{equation}\label{eqrns04}
e^{-i \mathbf{p}\cdot \mathbf{x}} = 4\pi \,\sum_{L,\,M} \,(-1)^L\,i^{L} \,j_{_L}(px)\, Y_{_{LM}}(\hat{\mathbf{p}})Y^*_{_{LM}}(\hat{\mathbf{x}}),
\end{equation}
where $j_{_L}(px)$ is the spherical Bessel function of order $L$. Therefore, the momentum space bound state wavefunction can be obtained:
\begin{equation}\label{eqbswf20}
\mathcal{U}_{E\kappa m} (\mathbf{p}) = 4\pi (-i)^l  \left( \begin{array}{l}
g_{_{E\kappa}}(p)\mathcal{Y}_{_{\kappa m}}(\hat{\mathbf{p}})\\
\\
f_{_{E\kappa}}(p)\boldsymbol \sigma \cdot \hat{\mathbf{p}} \mathcal{Y}_{_{\kappa m}}(\hat{\mathbf{p}})
\end{array}\right), 
\end{equation}
where the momentum space upper and lower component radial wavefunctions are defined as 
\begin{eqnarray}
\label{eqrns06a}
g_{_{E\kappa}}(p) = \int \mathrm{d}x \, x \,g_{_{E\kappa}}(x)\,j_{_{l}}(px), \,\,\,\,\,\,\,\,\,\,\,\,\,\,\,\,\,\,\,\,\\
\label{eqrns06b}
f_{_{E\kappa}}(p) = \mathrm{sgn}(\kappa)\int \mathrm{d}x \, x \,f_{_{E\kappa}}(x)\,j_{_{l'}}(px).
\end{eqnarray}

\begin{figure}[\h]
\centerline{\psfig{file=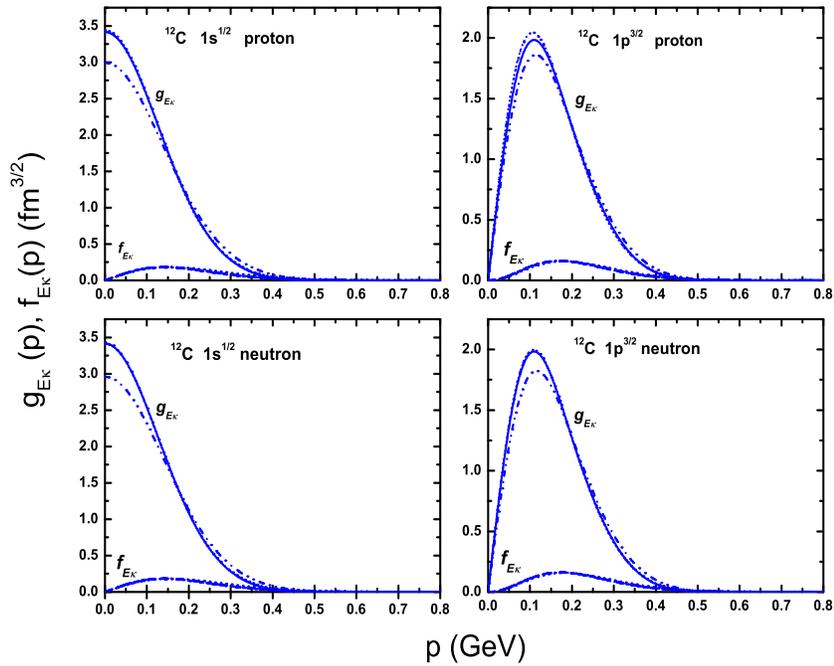,width=12.5cm, height=10.5cm}}
\vspace*{8pt}
\caption{The momentum space radial wavefunctions of the bound nucleon. The top(bottom) panel contains the curves for proton (neutron) bound to the 1s$^{1/2}$ and 1p$^{3/2}$ orbitals of $^{12}$C.  The solid (dashed), dotted(dash-dotted), dash-dot-dotted (short-dashed), and short-dotted (short-dash-dotted) curves of $g_{E\kappa}(f_{E\kappa})$ correspond to QHD-I, QHD-II, NL3, and FSUGold parameterizations, respectively, to the RMF model.}
\label{fig04}
\end{figure}

Figure \ref{fig04} shows the radial wavefunctions as a function of the magnitude of three-momentum for the nucleons in the 1s$^{1/2}$ and 1p$^{3/2}$ orbitals of $^{12}$C. These wavefunctions are obtained by using the RMF models mentioned earlier. The plots indicate that, under proper kinematic settings, one can explore the missing momentum distribution of the cross section and subsequently investigate how the nuclear medium effects influence the reaction process. In short, the characteristic curves of the radial components wavefunctions suggest that significantly larger cross sections can be obtained by carefully tuning the magnitude of the missing momentum in the range $\leq 0.3$ GeV. 

\section{Model for Nuclear Weak Transition}\label{sec03}

Apart from Refs.\cite{PRC.D12.2049.1974,APA48.293.1978,PRC.D18.2290.1978,PRC.D24.2369.1981,PRC82.025501.2010} there is very little theoretical work on neutrino-induced associated production of strange particles. To our knowledge, this paper represents the first attempt of a theoretical description of this process from nuclei. As such, we develop our model of the nuclear weak transition matrix element for $\rm KY$-production based on the photo- and electroproduction of mesons in association with baryons from nuclei.\cite{PRC55.318.1996,Nucl.Phys.A695.2001,PRC61.014604.1999,LJFU2000,PRC55.2571.1997} In doing so, we invoke the impulse approximation whereby one assumes that the incident neutrino interacts with a single bound (off-shell) nucleon inside nuclear medium via the exchange of a single weak gauge boson, and subsequently a pseudoscalar meson is produced in association with hyperon. The rest of the nucleons behave like spectators and make up the residual nucleus. This assumption leads to the embedding of the elementary hadronic weak current operator\cite{PRC82.025501.2010} inside the nuclear medium without changing its on-shell details.\cite{PRC55.2571.1997} Nuclear effects are taken into account using the RMF approximation to the Walecka model in the description of the bound state nucleon. 

In similar spirit to the above arguments the conventional interpretation of the Feynman diagram representing the vertex approximation in Fig. \ref{fig02} is necessary. The nuclear weak transition matrix element is constructed in terms of the the leptonic and nuclear currents, vector gauge boson propagator, and weak vertex coupling constants associated to the Feynman diagram that reinforces the assumptions made under the impulse approximation. That is,
\begin{equation}\label{eqnme01}
-i\mathcal M_{fi} =  \eta  L_{\mu} \dfrac{-i g^{\mu \nu}}{M^2_{_W}} H_{\nu}.
\end{equation}

The leptonic current $L_{\mu}$, is completely determined from electroweak theory: 
\begin{equation}
\label{eqnme02}
L_{\mu}  = \overline{\mathbf u}_l(\boldsymbol k',\,h') \gamma _{\mu} (I - \gamma_5) \boldsymbol \nu ( \boldsymbol k),
\end{equation}
where the incident neutrino and the outgoing lepton are described by the helicity representation of the Dirac spinor fields. The helicity representation of the Dirac spinor of the outgoing lepton can be written as 
\begin{equation}\label{eqnme02a}
\mathbf{u}_l(\boldsymbol k',\,h') = \sqrt{\dfrac{E_{k'}+M_{_l}}{2E_{k'}}}\left( \begin{array}{c}
\boldsymbol\phi^{h'}(\hat{\mathbf{k}}') \vspace*{6pt}\\ 
\frac{h' |\mathbf{k}'|}{E_{k'}+M_{_l} }\boldsymbol\phi^{h'}(\hat{\mathbf{k}}')
\end{array}\right). 
\end{equation}
This Dirac spinor is non-covariantly normalized and Pauli spinors, $\boldsymbol\phi^{h=\pm1}(\hat{\mathbf{k}}\equiv\hat{\mathbf{k}}(\theta,\,\phi))$, are defined as
\begin{equation}\label{eqnme02b}
\boldsymbol\phi^{h=+1}(\hat{\mathbf{k}}) = \left(\begin{array}{c}
\cos(\theta/2)\vspace*{6pt}\\
\sin(\theta/2)e^{i\phi}
\end{array} \right), \,\,\,\,\,\, \boldsymbol\phi^{h=-1}(\hat{\mathbf{k}}) = \left(\begin{array}{c}
 -\sin(\theta/2)e^{-i\phi}\vspace*{6pt}\\
\cos(\theta/2)
\end{array} \right).
\end{equation} 
Note however, that neutrinos are assumed to be massless, and subsequently they can only be described by the negative helicity states. In this paper we do not include Coulomb distortions for the CC processes, because of the fact that their effects on the reaction cross section are usually assumed to be negligible.

In the framework of the impulse approximation, the nuclear weak current is written as the sum of the contribution of the elementary hadronic weak currents of the associated production from the individual nucleons inside the nucleus:\cite{PRC73.024607.2006,nucl.th.0805.2344.2008}
\begin{equation}
\label{eqnme03}
{H}^{\mu} = \int \mathrm{d^3} \mathbf x  \,{\mathcal{J}}^{\mu}(\mathbf x)\,e^{i \mathbf{q \cdot x} },
\end{equation}
where $ {\mathcal{J}}^{\mu}(\mathbf x)  $ is the weak hadronic transition current for the associated production from a single nucleon:\cite{PRC51.2739.1995,Nucl.Phys.A623.1997} 
\begin{equation}\label{eqnme04}
\mathcal{J}^{\mu}(\mathbf x)  =  \overline{\boldsymbol \Psi}^{^{(-)}}_{\hat{\mathbf i}'_2, s'_2}(\mathbf x,\,\mathbf p'_2)\, \left[\boldsymbol \phi ^{^{(-)}}(\mathbf x,\,\mathbf p'_1)\right]^* {J}^{\mu}(q)\, \mathcal{U}_{_{E \kappa m}}(\mathbf x ),
\end{equation}
where ${J}^{\mu}(q)$ is the elementary hadronic weak current operator that has already been developed for the neutrino-induced strange particle production on a free (on-shell) nucleon.\cite{PRC82.025501.2010} $ {\boldsymbol \Psi}^{^{(-)}}_{\hat{\mathbf i}'_2, s'_2}(\mathbf x,\,\mathbf p'_2)$ and $\boldsymbol \phi^{^{(-)}} (\mathbf x,\,\mathbf p'_1)$ are the wavefunction of the final state hyperon and kaon, respectively, with the outgoing boundary conditions as indicated by the minus sign superscripts. At this stage, both of these wavefunctions are kept general. Note also that $\hat{\mathbf i}'_2$ is an arbitrary spin quantization axis in the rest frame of the hyperon and $s'_2$ is the spin projection of the hyperon. The weak coupling factor $\eta$, may take either of the following values depending on the production channel being studied:
\begin{equation}\label{eqnme05}
-\dfrac{\eta}{M^2_{_W}} = \dfrac{G_{_{F}}}{\sqrt{2}} \times \left\lbrace \begin{array}{ll}
\cos \theta _{_C}, & \text{for CC} \, \& \, \Delta \text S = 0 \vspace*{6.5pt}\\
\sin \theta _{_C}, & \text{for CC} \, \& \, \Delta \text S = 1 \vspace*{6.5pt}\\
\frac{1}{2}, & \text{for NC}, \\
\end{array}\right. 
\end{equation}
where  $G_{_F} = 1.166 \times 10^{^5}$ GeV$^{-2}$ is the Fermi constant, $\theta _{_{C}}$ is the phenomenological constant known as the Cabibbo angle ($\cos\theta_{_{C}} \approx 0.97$). Note that this formulation is valid in the limit $Q^2 \ll M^2_{_W}$ $_{•}(Q^2 = - q^2)$.

\subsection{Relativistic plane wave impulse approximation}

The description of nuclear scattering processes is always challenging due to nuclear effects which must be taken into account in order to establish a reasonable theory. Within the framework of the impulse approximation we assume that these effects arise when the underlying elementary process is embedded inside the nuclear medium. In section \ref{sec02} we have incorporated the basic effects in terms of the bound state wavefunction of the nucleon. There are also other nuclear effects that may come into play and their origin can be traced to the final state interactions (FSIs) like the strong nuclear force and/or the Coulomb interaction (for CC reactions). In this model, however, we assume that the effects form the FSIs are negligible and hence the outgoing kaon and hyperon are described by plane waves. 

In this work, we present a formalism based on the RPWIA. This is done for a number of reasons. First, it serves as a starting point by providing baseline calculations against which we can compare the full, if any, RDWIA calculations as well as the high-statistic data that will possibly be available in the near future. Second, the inclusion of full distorted waves represents a considerable numerical challenge as evidenced by similar work done for proton-induced reactions.\cite{PhysRevC.83.044607.2011,PhysRevC.83.044616.2011} It might therefore be very difficult to disentangle effects related to nuclear structure, medium effects, the description of the bound state wavefunction and the elementary process as well as distortions all in one go. Finally, despite their significance role in the present and future data analysis in the neutrino-oscillation experiments and the proton decay searches, the scarcity of theoretical work on the neutrino-induced strangeness associated production on nuclei, $A(\nu,\,lKY)B$, is also what has lead us to establish this baseline formalism. 

In so doing, we have resorted to using the consequences of a strong analogy between the studies of reaction processes that proceed via the weak interactions and that of their counterparts that proceed via the electromagnetic interactions. In Refs.\cite{PRC48.816.1993,PRC55.318.1996,Nucl.Phys.A695.2001}, in which the authors extensively investigated the $A(\gamma,\,\pi N)B$, $A(e,\,e'\pi N)B$, and $A(\gamma,\, KY)B$ processes within the RPWIA and RDWIA frameworks. What they found out was that the final state distortions only appear to suppress the overall magnitude of the RPWIA cross sections without causing any substantial change of their shapes. Another example is in Refs.\cite{PRC69.035501.2004,PRC73.025501.2006} where a RPWIA formalism was used to investigate the strange quark effects of the axial form-factor of the nucleon via neutrino-nucleus scattering. In what follows, therefore, we evaluate the invariant matrix element in the RPWIA framework. 

In the plane wave limit the kaon and hyperon are described by the wavefunctions of the forms: $e^{i \mathbf p'_1 \cdot \mathbf x}$ and  $\mathbf U_{_Y}(\mathbf p'_2, \hat{\mathbf i}'_2, s'_2)e^{i\mathbf p'_2 \cdot \mathbf x} $, respectively, and both are on their respective mass shell.  Thus in the RPWIA approximation, the nuclear weak transition, which is proportional to the Fourier transform of the bound state wavefunction of the struck nucleon, can be written as
\begin{equation} \label{eqpwia01}
{H}^{\mu} = \overline{\mathbf U}_{_Y}(\mathbf p'_2,\,\hat{\mathbf i }'_2,\,s'_2 )\,{J}^{\mu}(q)\,\mathcal{U}_{_{E \kappa m}}(\mathbf p_m),
\end{equation}
where $\mathbf{p}_m = \mathbf p'_1 + \mathbf p'_2 - \mathbf q$ and $\mathcal{U}_{_{E \kappa m}}(\mathbf p_m)$ in the momentum space wavefunction for bound nucleon as defined in Eq. (\ref{eqrns03}) and its explicit form is also given in Eq. (\ref{eqbswf20}). The weak hadronic current operator, ${J}^{\mu}(q)$, is briefly presented in section \ref{sec04}. In the plane wave limit the outgoing hyperon is described by the non-covariantly normalized Dirac spinor:
\begin{equation}\label{eqpwia01b}
\mathbf{U}_Y(\mathbf{ p}'_2, \hat{\mathbf{ i} }'_2, s'_2) = \sqrt{\frac{E_{p'_2} + M_{_Y}}{2E_{p'_2}}} \left( \begin{array}{c}
\boldsymbol\phi^{s'_2}(\,\hat{\mathbf i }'_2) \vspace*{6pt}\\ 
\frac{\boldsymbol \sigma \cdot \mathbf{p}'_2}{E_{p'_2} + M_{_Y}} \boldsymbol\phi^{s'_2}(\,\hat{\mathbf i }'_2)
\end{array}\right),  
\end{equation}
where the generalized Pauli spinor with respect to any arbitrary spin quantization axis, $\hat{\mathbf i } \equiv \hat{\mathbf i }(\alpha,\,\beta)$, is defined in terms of the Wigner D-function as
\begin{equation}\label{eqpwia01c}
\boldsymbol\phi^{s}(\alpha, \beta )= \sum_{s'=\pm \frac{1}{2}} \mathcal{D}^{^{\frac{1}{2}}}_{s s'}(\alpha, \beta, 0)\,\boldsymbol\chi^{^{\frac{1}{2}}}_{s'}.
\end{equation}

For the neutrino-induced CC process the outgoing lepton would be a negatively charged muon with mass, and hence that makes it detectable by the modern detectors. The same can be said for the outgoing $\mathrm K^{^+}$-meson. As for the final state hyperon, it is only the direction of $\mathbf p'_2$ which is needed to be tracked regardless of its magnitude. In contrary, the recoiling residual nucleus goes undetected. Therefore, by referring back to Eq. (\ref{eqdxs01}), we integrate over the recoil three-momentum of the residual nucleus, $\mathbf P'$, and the outgoing hyperon energy, $E_{p'_2}$, and then the five-fold differential cross section can be obtained:
\begin{equation}\label{eqres01}
\dfrac{\mathrm d^5 \sigma}{\mathrm d E_{k'}\mathrm d(\cos \theta_l)\mathrm d E_{p'_1}\mathrm d\Omega_{p'_1} \mathrm d \Omega_{p'_2} } =   \dfrac{1}{2(2\pi)^7} \dfrac{E_k' E_{p'_2} E'\, |\mathbf{k}'|\, |\mathbf{p}'_1|\, |\mathbf{p}'_2|}{\left| E' + E_{p'_2}(1 - \mathbf{p}'_2 \cdot (\mathbf{q} - \mathbf{p}'_1)/|\mathbf p'_2|^2)\right|} \,|\mathcal{M}_{fi}|^2,   
\end{equation}
where we have imposed non-covariant normalizations. The norm squared invariant matrix element is derived from Eq. (\ref{eqnme01}). For the $\rm KY$-production channel that proceeds via the CC and $\Delta S = 0$ process:
\begin{equation}\label{eqres02b}
|\mathcal{M}_{fi}|^2_{_{CC,\,\Delta S=0}} = \dfrac{G^2_F}{2}\cos^2 \theta_{_C} \sum_{h'}[L_{\mu}L^{\dagger}_{\nu}]\sum_{m,\,s'_2}[H^{\mu}{H^{\nu}}^{\dagger}],
\end{equation}
where $L_{\mu}$ and $H_{\mu}$ are defined in Eqs. (\ref{eqnme02}) and (\ref{eqpwia01}). In this paper we present results for the unpolarized differential cross sections by summing over the initial nucleus states, helicity states of final lepton, and spin states of hyperon.

\section{Elementary Hadronic Amplitudes} \label{sec04}

The model-independent hadronic weak current operator for the neutrino-induced strange particle production from on-shell nucleon is discussed in this section. Note that, in this work, we take the same underlying elementary process and embed it unmodified inside the nuclear medium (i.e., the impulse approximation). The elementary production process of interest is of the form:
\begin{equation}\label{eqel01}
\nu(k) + N (p) \rightarrow l(k') + K(p'_1) + Y(p'_2).
\end{equation}

The hadronic vertex is represented by the generalized current operator which may be constructed based on the global structure the operator processes, i.e., the current operator is a Lorentz vector and $4 \times 4$ matrix. Having sandwiched the current operator between two on-shell Dirac spinors, we can invoke the linear decomposition in the Dirac spinor space in terms of the bilinear covariant basis:\cite{RQM-BjDr-1964,QED.WGJR.113.2009} $I, \, \gamma^5,\, \gamma^{\mu},\,\gamma^{5} \gamma^{\mu}$,\, $\sigma^{\mu \nu}$. In doing so, we introduce five unknown vertex amplitudes which demand further expansion by using three independent four momenta, the metric tensor, and the Levi-Cevita tensor:  $q^{\mu}$, $ p^{\mu}$, ${p'_1}^{\mu}$,  $ g^{\mu \nu} $, and $\varepsilon^{\mu \nu \alpha \beta}$, which we have at our disposal. By employing the Dirac algebra and Gordon-like identities on the gamma matrices and four vector momenta, the most general representation of the weak hadronic current operator takes the following form:

\begin{equation}\label{eqc4_18}
\begin{array}{ll}
J^{\mu}(q) = & \tilde{A}^{\mu}I + \tilde{B}^{\mu}\gamma_{5} + \tilde{C}_1 \gamma^{\mu} + \tilde{C}^{\mu} \displaystyle{\not} q + \tilde{D}_1 \gamma_5 \gamma^{\mu}\vspace*{0.15cm}\\
& + \tilde{D}^{\mu} \gamma_5 \displaystyle {\not} q + \tilde{D}_5 \displaystyle{\not}q \gamma^{\mu} + \tilde{D}_6 \gamma_5 \displaystyle {\not} q \gamma^{\mu},
\end{array}  
\end{equation}
where
\begin{equation}\label{eqc4_19}
\begin{array}{ll}
\tilde{A}^{\mu} & = \tilde{A}_1 q^{\mu} + \tilde{A}_2 p^{\mu} + \tilde{A}_3 {p'_2}^{\mu} + \tilde{A}_4 \varepsilon^{\mu \nu \alpha \beta}q_{\nu}{p}_{\alpha}{p'_2}_{\beta},\vspace*{0.15cm}\\
\tilde{B}^{\mu} & = \tilde{B}_1 q^{\mu} + \tilde{B}_2 p^{\mu} + \tilde{B}_3 {p'_2}^{\mu} + \tilde{B}_4 \varepsilon^{\mu \nu \alpha \beta}q_{\nu}{p}_{\alpha}{p'_2}_{\beta},\vspace*{0.15cm}\\
\tilde{C}^{\mu} & = \tilde{C}_2 q^{\mu} + \tilde{C}_3 p^{\mu} + \tilde{C}_4 {p'_2}^{\mu},\vspace*{0.15cm}\\
\tilde{D}^{\mu} & = \tilde{D}_2 q^{\mu} + \tilde{D}_3 p^{\mu} + \tilde{D}_4 {p'_2}^{\mu}.
\end{array}
\end{equation}

As similar procedure has already established two such form factors for the two-body electromagnetic interactions of Dirac particles. These two independent form factors have become a successful means in the study of the structure of nucleons in detail by using electron scattering and able to interpret the experimental data reasonably well. Our method is also similar in spirit with the CGLN mechanism developed for the photo- and electroproduction of mesons on nucleons and nuclei.\cite{PR106.1345.1957} It is also worth noting that, unlike the electromagnetic production processes, here we do not need to impose symmetries such as parity, time-reversal, and charge conjugation, since the weak interaction is known to violate each of them.  

\subsection{Born term approximation}

The model-independent representation of the weak hadronic current operator for $\rm KY$-production consists of eighteen unknown parametrization amplitudes.\footnote{The detailed derivation of the most general hadronic weak current operator and the extraction of these amplitudes from the Born diagrams for CC $\rm KY$-production processes are given in Ref.\cite{PRC82.025501.2010}.} However, one needs to resort to model-dependent calculations to determine them. In doing so, we employ the Born term approximation, in which the production vertex is approximated by a series of the Feynman tree-level diagrams, to describe the invariant amplitudes in terms of the hadronic weak form factors and Born coupling constants. In this approximation we only consider background contributions: $s$-channel via nucleon exchange, $u$-channel via ground-state hyperon exchange, and $t$-channel via kaon exchange. Note that the labels $s$, $u$, and $t$ are the Lorentz invariant kinematic variables:
\begin{equation}\label{eqbt01}
s = (q + p)^2, \,\,\,\,\,\,\,\,\,\,t = (q - p'_1)^2,\,\,\,\,\,\,\,\,\,\,u = (q - p'_2)^2.
\end{equation}

The Born diagrams, in general, characterized by three different coupling vertices. First, we have a vertex for a single baryon weak transition within the SU(3) baryonic octet. Due to this transition three standard form factors: $F_1(Q)$, $F_2(Q)$, and $G_{_A}(Q)$ come into picture. According to Ref.\cite{PRL10.531.1963} the vector form factors are determined from the corresponding electromagnetic form factors by using the CVC  hypothesis, whereas we adopt the method used by Ref.\cite{PRD74.053009.2006} in order to determine the axial form factor. The explicit forms of these standard form factors for the case of the weak CC transitions amongst the SU(3) baryon octet states are summarized in Table IV of Ref.\cite{PRC82.025501.2010}. 

Second, we have the weak transition vertex of a pseudoscalar meson that primarily comes through the $t$-channel diagram. This vertex introduces either of the following pseudoscalar form factors: $F_{K^{^+}}(q)$, $F_{K^{^{+,0}}}(q)$, or $F_{K^{^{0,+}}}(q)$. We make use of the method employed in Ref.\cite{APA48.293.1978} whereby the author determined their values based on their relationship with the vector meson form factors. Finally, we have the strong pseudoscalar coupling vertex, that introduces the strong coupling constants,  $g_{_{KYN}}$, into the formalism of the process and their values are taken from the broken SU(3) symmetry predictions based on the well-known phenomenological coupling constant $g_{_{\pi NN}}$.\cite{PR151.1322.1966,PRC.51.R1074.1995,Nucl.Phys.A691.2001} 

It is worth mentioning significant uncertainties that our relativistic formalism may suffer from. The main uncertainty is presented by the estimated strong coupling constants used in our calculation owing to the fact that their exact values have never been determined to date. The Born approximation is also another source of ambiguity due to the lack of improved understanding as to how and which resonance and/or background(non-resonance) terms justifiably contribute to the reaction channels of great interest. There are other uncertainties such as non-local and off-shell ambiguities incurred by the impulse approximation, specially, as one attempts to embed elementary weak hadronic current amplitudes inside nuclear medium in which the nucleons are bound. In addition, plane wave limit treatment of the final state kaon and hyperon is also a naive approximation in a sense that the exclusive cross sections could be sensitive to the distortion effects coming from the FSIs. In a future these issues must be addressed in order to give a complete theoretical picture of strangeness associated production via the neutrino-nucleus weak interaction. 

\section{Results and Discussion} \label{sec05}

In order to generate numerical results we have chosen a $^{12}$C target and particularly investigated the $\mathrm K^{^+}\Lambda$-production channel from the bound neutron that proceeds via the CC and $\Delta S = 0$ process. Note however, that our formalism can easily be generalized to other nuclei such as $^4$He, $^{16}$O, $^{40}$Ca, and $^{208}$Pb and production channels such as K$^{^+} \Sigma^{^+}$, K$^{^+} \Sigma^{^0}$, and K$^{^0} \Sigma^{^+}$. The most general form of the differential cross section is presented in Eq. (\ref{eqdxs01}).  Nevertheless, one needs to eliminate the four-dimensional $\delta$-function which is incorporated to enforce that the overall energy-momentum conservation at the nuclear interaction vertex is preserved. Upon eliminating the $\delta$-function the five-fold differential cross section is obtained in  Eq. (\ref{eqres01}), and in what follows this quantity will be used to investigate the neutrino-induced associated production process in detail.  

\begin{figure}[th]
\centerline{\psfig{file=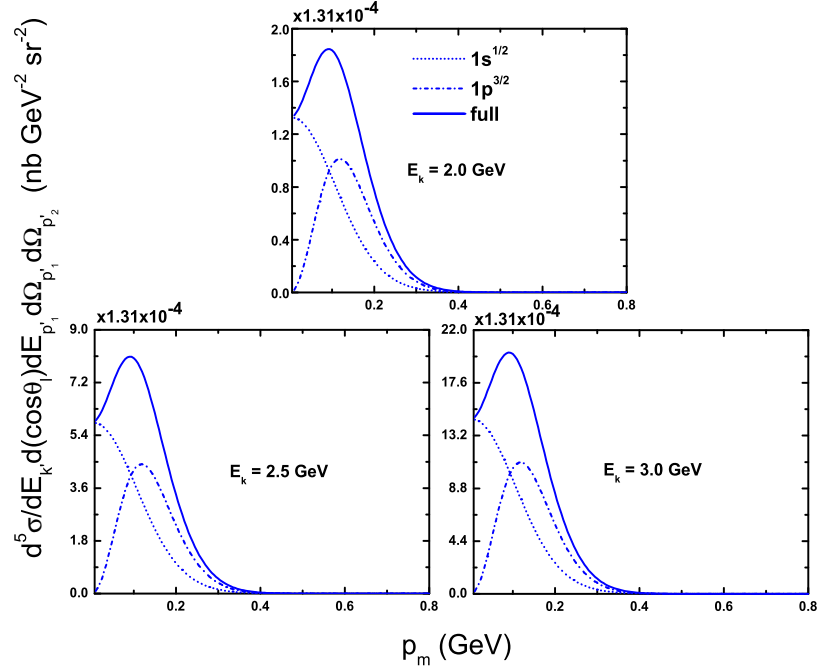,width=12cm, height=10.25cm}}
\vspace*{8pt}
\caption{Missing momentum distribution of the differential cross section for the quasifree $\rm K^{^+}\Lambda$-production. The short-dashed (short-dash-dotted) line is the contribution from the neutrons in $1s^{1/2}(1p^{3/2})$ orbital of $^{12}$C and the solid line is their total sum. The kinematics inputs are fixed at $\omega = 1.46$ GeV, $Q^2 = 0.05$ GeV$^2$, and $\theta'_1 = 10^{^0}$.}
\label{fig05}
\end{figure}

In this paper we present the unpolarized cross section by summing over spin states. Note however, that our theoretical framework is also suitable to investigate polarization observables. As for the spin quantization axis, $\hat{\mathbf{i}}'_2$, it is fixed along the direction of the three-momentum of the outgoing hyperon (i.e., $\hat{\mathbf{i}}'_2 \equiv \hat{\mathbf{i}}'_2(\theta'_2,\,\phi'_2)$). This particular choice is to keep the consistency of our description with regard to the weak interaction naturally favoring particles with longitudinal spin polarization ( i.e., along the direction of its three-momentum), and even the ones with negative helicity states in particular to account for the violation of fundamental symmetries such as parity by the weak interaction.

The three-momenta of the outgoing K-meson and $\Lambda$-hyperon are set to lie on the same plane ( and for simplicity sake we call it production plane) by imposing the constraint: $\phi'_1 = \phi'_2 - 180^{^0} = \phi$. Furthermore, we limit ourselves to the coplanar setup whereby the scattering and production planes coincide (i.e., $\phi = 0^{^0}$), since this geometrical setup, as pointed out in the study of photoproduction of associated production in Ref.\cite{Nucl.Phys.A695.2001},  generates larger cross sections relative to the out-of-plane setups. 

\begin{figure}[th]
\centerline{\psfig{file=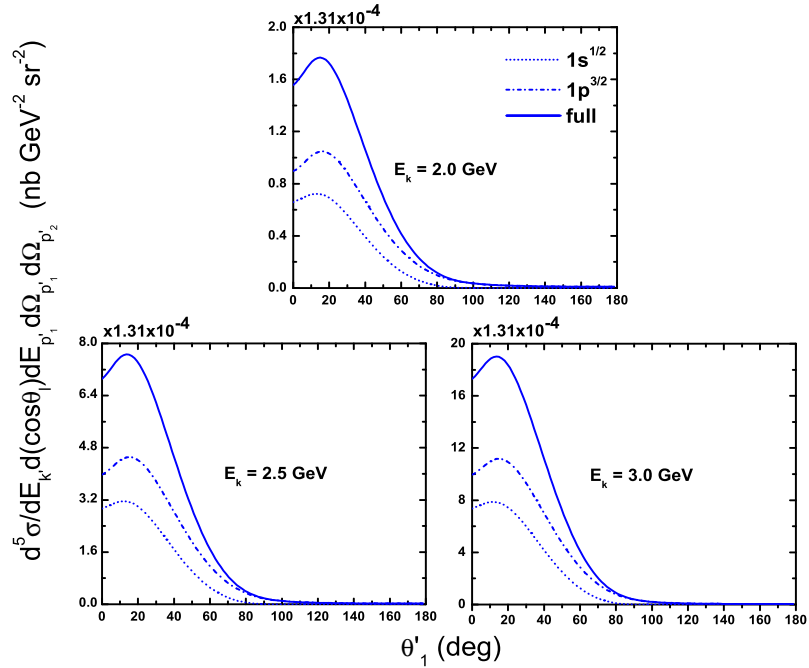,width=12cm, height=10.25cm}}
\vspace*{8pt}
\caption{Differential cross section as a function of kaon angle for the quasifree $\rm K^{^+}\Lambda$-production from CC neutrino-nucleus scattering in the intermediate energy region. The kinematics inputs are fixed at $\omega = 1.46$ GeV, $Q^2 = 0.05$ GeV$^2$, and $p_m = 0.12$ GeV. The short-dashed (short-dash-dotted) line is the contribution from the neutrons in $1s^{1/2}(1p^{3/2})$ orbital of $^{12}$C and the solid line is their total contribution.}
\label{fig06}
\end{figure}
	
It is worthy noting that, the theoretical framework of the associated production of strange particles on nuclei presents kinematic flexibility in comparison to the underlying production on unbound nucleon. In the coplanar setup, we adopt the kinematics settings used in Ref.\cite{PRC55.318.1996,Nucl.Phys.A695.2001}. First, we implement the setting whereby $E_k$, $\omega$, $Q^2$, $\theta'_1$, and $p_m$ are specified as input variables. Note that we used ${p}_m$ in a place of $|\mathbf{p}_m|$. In this setting one needs to solve for $T_{_K}$, $T_{_Y}$, and $\theta'_2$. For convenience we call this arrangement as ``setting I". Second, we make use of the kinematic setting that appears to be more natural given the expression of the five-fold differential cross section. In this setting $E_k$, $\omega$, $Q^2$,  $T_{_K}$, $\theta'_1$, and $\theta'_2$ are fixed beforehand as input variables and then one has to determine $\mathbf p_m$ and $T_{_Y}$ by solving the overall four-momentum conservation. This arrangement is labeled as ``setting II". Under setting II, $p_m$ is set to vary in a wider range, and hence we will be able to probe the momentum distribution of the bound nucleon. 

\begin{figure}[th]
\centerline{\psfig{file=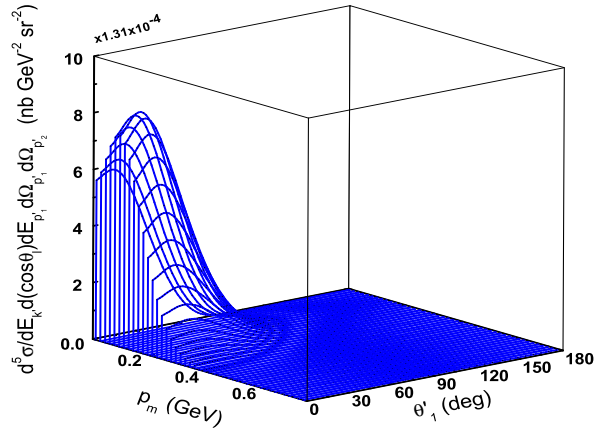,width=0.725\columnwidth, height=0.565\columnwidth}}
\vspace*{8pt}
\caption{Total differential cross section of quasifree $\rm K^{^+}\Lambda$-production from CC neutrino-nucleus scattering as a function of the missing momentum and outgoing kaon angle at $E_k = 2.5$ GeV, $\omega = 1.46$ GeV, and $Q^2 = 0.05$ GeV$^2$.}
\label{fig07}
\end{figure}

\subsection{Results of setting I}

This kinematics setting is more like that of the associated production on free nucleon. Here we investigate the missing momentum and kaon angle distributions of the differential cross section for neutrino energies $E_k = 1.5,\,\,2.0,\,\,3.0$ GeV within the plane wave impulse approximation. Under the quasifree treatment we fix  $\omega$ at $1.46$ GeV and  $Q^2$ at $0.05$ GeV$^2$ throughout this section. In Fig. \ref{fig05} we display the missing momentum distribution of the differential cross section of neutrino-induced associated $\rm K^{^+}\Lambda$-production on $^{12}$C at $\theta'_1 = 10^{^0}$. The graphs illustrate the contributions from neutrons in the $1s^{1/2}$ and $1p^{3/2}$ orbitals of $^{12}$C and their sum for the $\rm K^{^+}\Lambda$-production. The results for the $^{12}$C orbitals clearly justifies the proportionality between the cross section and the bound state wave function. The contribution from the $1s^{1/2}$ dominates over the $1p^{3/2}$ orbital for the missing momentum in the range $p_m < 0.1 $ GeV, where as for the rest of the range it is the other way round that holds otherwise (up to the overall scaling factor).

\begin{figure}[th]
\centerline{\psfig{file=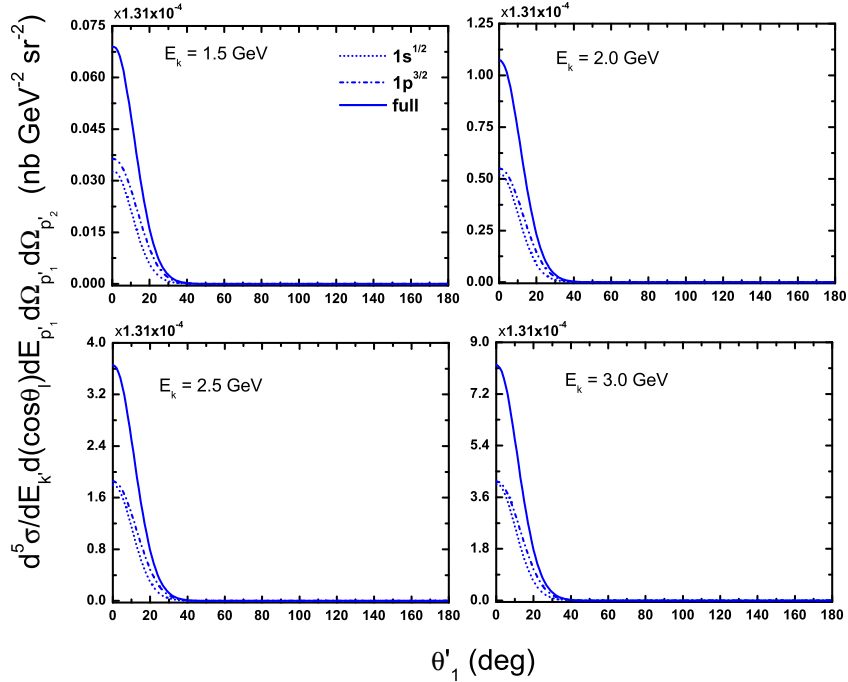,width=1.0\columnwidth}}
\vspace*{8pt}
\caption{Koan angular distribution of the differential cross section for neutrino-induced CC associated $\rm K^{^+}\Lambda$-production on $^{12}$C in the intermediate energy region under kinematics setting II . The curves correspond to the individual contributions of the orbitals and their sum  at $\omega = 1.32$ GeV, $Q^2 = 0.12$ GeV$^2$, $T_{_K} = 300 $ MeV, and $\theta'_2 = 0.5^{^0}$.}
\label{fig08}
\end{figure}

In Fig. \ref{fig06} we show the quasifree differential cross section as a function of the kaon angle and this is done by specifying the magnitude of the missing momentum at $p_m = 0.12$ GeV. The $1p^{3/2}$ contribution clearly dominates for the whole range of kaon angle and the curves indicate that the peak occurs at foreword angle. Moreover, the curves of the cross section level off slowly suggesting that for the wide range of forward angle it is highly likely to detect the outgoing K-meson produced in association with $\Lambda$-hyperon on bound neutron struck by incident neutrino with intermediate energy. It is important to note that the impulse approximation is also more reliable for kaon produced in the direction of forward angles.\cite{PRC70.044008.2004,IJMPE19.2369.2010}

\begin{figure}[th]
\centerline{\psfig{file=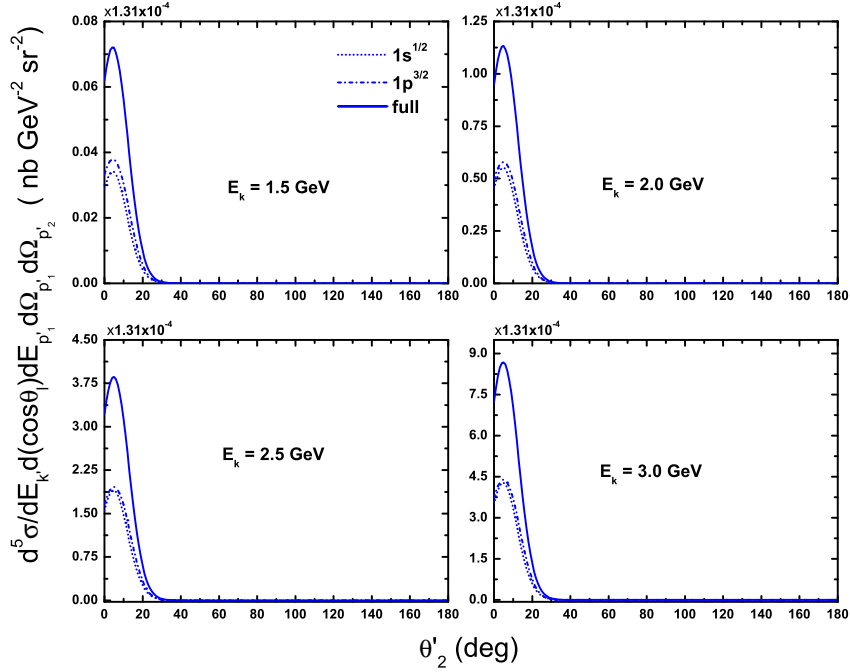,width=1.0\columnwidth}}
\vspace*{8pt}
\caption{Angular distribution of the differential cross section with respect to the $\Lambda$-hyperon angle for neutrino-induced CC $\rm K^{^+}\Lambda$-production on bound neutron in $^{12}$C nucleus in RPWIA framework under kinematic setting II at $\theta'_1 = 5^{^0}$.}
\label{fig09}
\end{figure}

Figure \ref{fig07} displays the three-dimensional angular and missing momentum distribution of the differential cross section.  This figure provides the wide coverage on the characteristic of the differential cross section for the quasifree production of strange particles on nuclei. We notice that the larger differential cross section can be obtained by tuning the missing momentum in the range $p_m \le 0.15$ GeV and the kaon angle in the range $\theta'_1 \le 30^{^0}$. The phase space factor suppresses the occurrence of the peak at small forward angles and causes the shift in the position of the peak as opposed to the dynamics factor of the cross section.  Nevertheless, both dynamics and phase space factors play more or less equal role in determining the shape of the cross section for the wider range of the kaon angle.

\subsection{Results of setting II}\label{sec041}

Numerical results for the Born term approximation in the plane wave limit are presented for the intermediate neutrino energies $E_k = 1.5,\,\,2.0,\,\,2.5,\,\,3.0$ GeV. The kinematics input values \,$\omega = 1.32$ GeV and $Q^2 = 0.12$ GeV$^2$ are maintained throughout the discussion of setting II results. Under this setting we will extensively explore the angular distributions with respect to kaon and hyperon angels as well as energy distributions with respect to kinetic energies of the outgoing kaon and muon.

\begin{figure}[th]
\centerline{\psfig{file=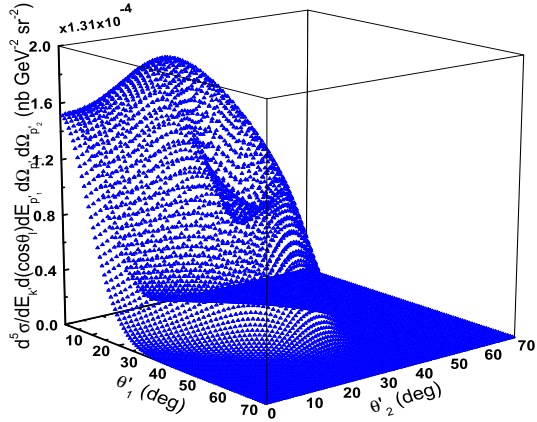,width=0.725\columnwidth, height=0.565\columnwidth}}
\vspace*{8pt}
\caption{Differential cross section as a function of kaon and $\Lambda$-hyperon angles at $E_k = 2.0$ GeV, $\omega = 1.32$ GeV, $Q^2 = 0.12$ GeV$^2$, and $T_{_K} = 350$ MeV.}
\label{fig10}
\end{figure}

In Fig. \ref{fig08} We show the kaon angular distribution for the CC $\rm K^{^+} \Lambda$-production at $T_{_K} = 300 $ MeV and $\theta'_2 = 0.5^{^0}$. By implementing the NL3 parametrization scheme to a single-particle relativistic shell model, we display the individual contributions of the neutrons in the $1s^{1/2}$  and $1p^{3/2}$  orbitals of $^{12}$C together with their sum for the $\rm K^{^+} \Lambda$-production. The contributions from the two orbitals are nearly the same up to some scaling factor. The distribution curves are peaked in the forward directions and they level off faster than the curves for the quasifree production. As pointed out in the study of CC neutrino-nucleus scattering,\cite{PRC68.048501.2003} the shapes and peaks of the cross sections mirror the missing momentum distribution of the stuck nucleon inside the nucleus. 

\begin{figure}[\h]
\centering
\includegraphics[width=1.0\columnwidth]{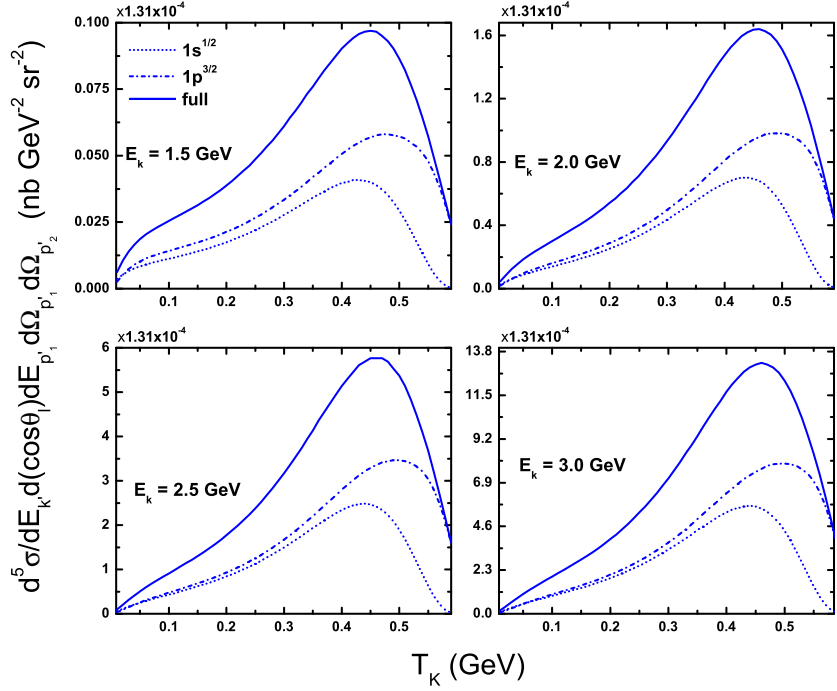} 
\caption{Kaon energy distribution of the differential cross section for the process $^{12}$C$(\nu,\,\mu^-K^{^+}\Lambda)$ under kinematic setting II at $\omega = 1.32$ Gev, $Q^2 = 0.12$ GeV$^2$,  $\theta'_1 = 10^{^0}$, and $\theta'_2 = 2.5^{^0}$.}
\label{fig11}
\end{figure} 

Similarly, in Fig. \ref{fig09} we display the $\Lambda$-hyperon angular distribution of the differential cross section for neutrino-induced associated production of $\rm K^{^+}\Lambda$ on the bound neutron in $^{12}$C target at $T_{_K} = 300$ MeV and  $\theta'_1 = 5^{^0}$. The cross sections from both orbitals attain their peak at forward $\Lambda$-hyperon angle indicating the similarity that the kaon and hyperon angular distributions may have in common. Moreover, both angular distributions fall off sharply before reaching $40^{^0}$. Furthermore, by tuning the  kinetic energy and angle of the koan, we are able to shift the peak and even obtain a double-peaked $\Lambda$-hyperon angle distributions.

\begin{figure}[th]
\centerline{\psfig{file=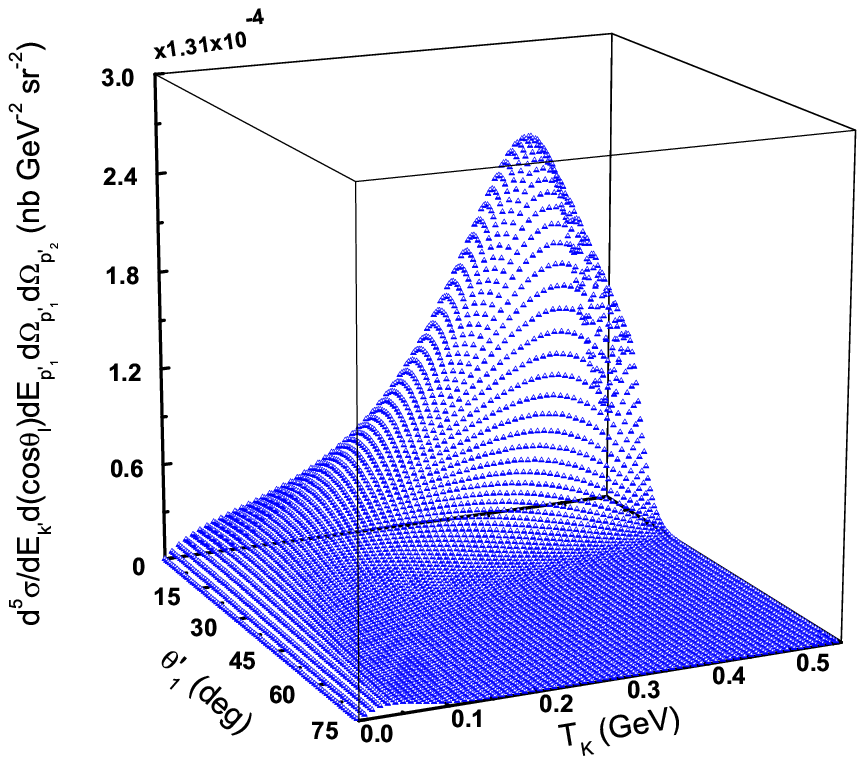,width=0.75\columnwidth, height=0.565\columnwidth}}
\vspace*{8pt}
\caption{Differential cross section as a function of kaon angle and kinetic energy. The kinematic setting II  inputs are fixed at $E_k = 2.0$ GeV, $\omega = 1.32$ GeV, $Q^2 = 0.12$ GeV$^2$, and $\theta'_2 = 2.5^{^0}$.}
\label{fig12}
\end{figure}

\begin{figure}[th]
\centerline{\psfig{file=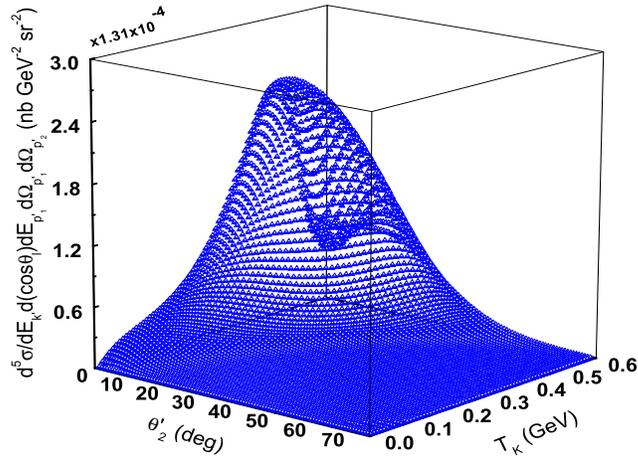,width=0.75\columnwidth, height=0.565\columnwidth}}
\vspace*{8pt}
\caption{Differential cross section as a function of $\Lambda$-hyperon angle and kaon kinetic energy. The kinematic setting II inputs are fixed at $E_k = 2.0$ GeV, $\omega = 1.32$ GeV, $Q^2 = 0.12$ GeV$^2$, and $\theta'_1 = 2.5^{^0}$.}
\label{fig13}
\end{figure}

In order to get a broader insight of our RPWIA calculations under kinematic setting II, three-dimensional plots of the differential cross sections are presented in Fig. \ref{fig10}, \ref{fig12}, and \ref{fig13}. These plots represent the total contributions of the $^{12}$C orbitals for the CC associated production. Figure \ref{fig10} displays the three-dimensional angular distribution as a function of the kaon and hyperon angles at $E_k = 2.0$ GeV, $\omega = 1.32$ GeV, $Q^2 = 0.12$ GeV$^2$, and $T_{_K} = 350$ MeV. We observe that at fixed value of $T_{_K}$ the peculiar shapes and peaks of the cross section are, in fact, in response to the missing momentum distribution of the struck neutron in $1p^{3/2}$ orbital. 

The richness of kinematics setting II can also be used to shed some light on the energy distributions of the differential cross section. Figure \ref{fig11} shows the kaon kinetic energy distribution of $\rm K^{^+}\Lambda$-production on bound neutrons in the $^{12}$C orbitals at $\theta'_1 = 10^{^0}$ and $\theta'_2 = 2.5^{^0}$. Upon fixing the angles of the outgoing koan and hyperon at small forward angles, the kaon energy distribution peaks from the $^{12}$C orbitals occur in the range $400 \le T_{_K} \le 500$ MeV. In Fig. \ref{fig12} we show the differential cross section as a function of kaon angle and kinetic energy at $\theta'_2 = 2.5^{^0}$. Furthermore, Fig. \ref{fig13} displays the differential cross section as a function of the $\Lambda$-hyperon angle and kaon kinetic energy at $\theta'_1 = 2.5^{^0}$. These three-dimensional plots provide a broader insight for the future experiments to probe the proper kinematics ranges that increase the likelihood of detecting the outgoing kaon and hyperon without interacting with the residual nucleus. 

\begin{figure}[th]
\centerline{\psfig{file=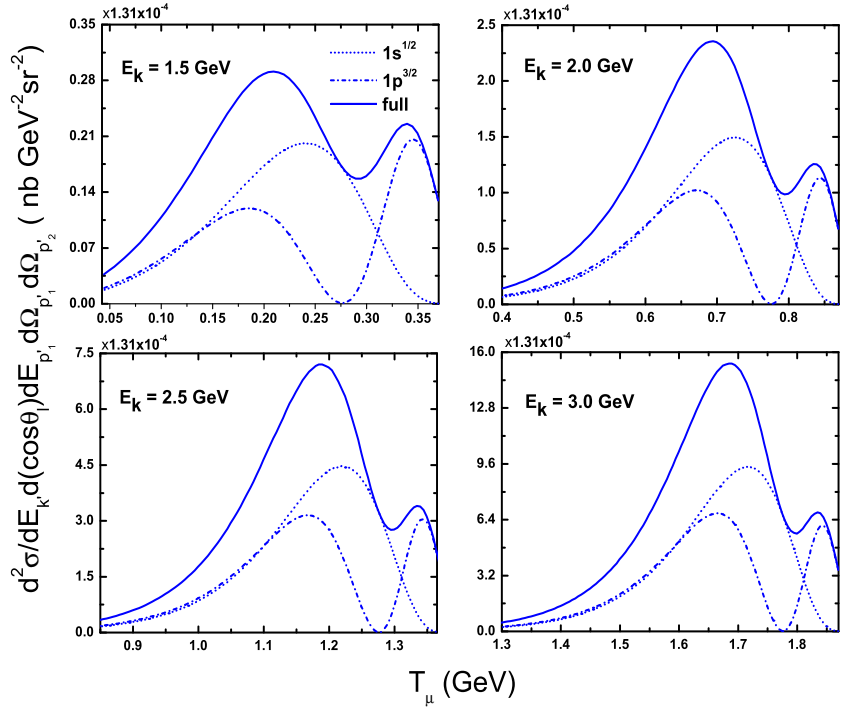,width=1.0\columnwidth, height=0.85\columnwidth}}
\vspace*{8pt}
\caption{Muon energy distribution of the differential cross section for $^{12}$C$(\nu,\,\mu^-K^{^+}\Lambda)$ under kinematics setting II in the intermediate energy region. The input variables are fixed at $Q^2 = 0.12$ GeV$^2$, $T_{_K} = 300$ MeV, and $\theta'_1 = \theta'_2 = 2.5^{^0}$.}
\label{fig14}
\end{figure}

Finally, in Fig. \ref{fig14} we show the muon kinetic energy distribution of the differential cross section at $T_{_K} = 300$ MeV and $\theta'_1 = \theta'_2 = 2.5^{^0}$. The double-peaked structure of the overall differential cross section can be traced back to the missing momentum distribution of the $p$-shell nucleons of $^{12}$C. We observe that as the muon kinetic energy increases, the $1s^{1/2}$ contribution rapidly  levels off and hence leaving the $1p^{3/2}$ orbital as the only candidate to extract the interaction information of neutrino-nucleus scattering. Note that, a similar remark can also be made for the kaon kinetic energy distribution of the differential cross section shown in Fig. \ref{fig11}.

\section{Summary and Conclusion} \label{sec06}

We have successfully presented the fully relativistic formalism of the neutrino-induced associated production of strange particles on nuclei within the plane wave impulse approximation framework. Under this approximation we have assumed that the incident neutrino scatters off a single bound nucleon and the remaining nucleons are treated as spectators. The nuclear medium effects such as Fermi motion, Pauli blocking, and nuclear binding are naturally incorporated via the nucleon bound state wavefunctions obtained from the improved effective Lagrangian of the Walecka model that makes use of the NL3 set of parameters.\cite{PRC55.540.1997} 

The leptonic weak current is completely derived in terms of the standard electroweak theory. However, the same does not apply for the nuclear weak current case owing to its complex structure. Therefore, after invoking the impulse approximation at the interaction vertex, the nuclear current is treated as the sum of the hadronic weak currents, which were developed for the elementary associated strangeness production on free nucleon.\cite{PRC82.025501.2010}   However, with in our formalism, this approximation presents some uncertainties such as off-shell ambiguity, which needs to be resolved in a future. In our numerical calculations we have employed the Born term approximation at the hadronic interaction vertex. Upon a systematic mapping, the eighteen parametrization amplitudes of the hadronic current operator are determined in terms of the standard vector-axial form factors and strong coupling constants.  

We have shown and discussed our results of the RPWIA calculations under different kinematic settings in the rest frame of the target nucleus. We investigated the angular and energy distributions of the five-fold differential cross section for the CC $\rm K^{^+} \Lambda$-production  on bound neutron in $^{12}$C  at the intermediate energies. The angular distributions peaked at forward angles under both kinematics settings. Under the first kinematic setting the missing momentum distribution clearly indicates the qualitative influence of the bound state wavefunctions have over the $1s^{1/2}$ and $1p^{3/2}$ contributions. On the other hand, the angular distribution shows the dominance of the $p$-shell for whole range of kaon angle. This setting does not provide a wide kinematic ranges from which the input variables are chosen. Nevertheless, we believe that it is a very useful setting in which one can extensively study the structure of electroweak interaction and even test the validity of various estimations of strong coupling constants.

The second kinematic setting leaves the missing momentum to vary freely, and as a consequence the angular and energy distributions clearly mimic the missing momentum distribution of the bound nucleon in the nucleus. This setting offers a freedom of tuning the kinematic inputs in the wider ranges and hence causing the substantial alteration to the shape and magnitude of the cross sections. Moreover, by choosing the input set that favors one orbital contribution over the other, one can separately study dynamics of the orbitals. We also believe that this kinematics flexibility can offer a wide ground in which one can give a reliable interpretation to the experimental data that will be made available in the near future.

In conclusion, The great merits of the RPWIA formalism are the following: (i) it provides a model which incorporates a number of important aspects of the reaction, (ii) it provides a first-order indication of the behavior of the cross sections for these types of reactions, (iii) from a theoretical perspective it allows a measure of analytical manipulation which greatly facilitates the numerical implementation and (iv) it will provide the baseline calculations for any possible future model which aims to incorporate final-state interactions. Moreover, in the Born approximation of the weak hadronic vertex, the inclusion of the relevant resonance contributions in the evaluation of the elementary matrix element at the intermediate energy region also enhances the importance of our study. One can also take into account the Coulomb distortion of the scattered muon in the CC associated production reactions. Our current calculations are timely in a sense that the ongoing neutrino-oscillation experiments such as MiniBooNE, K2K, and MINER$\nu$A, as well as other experiments dedicated for proton decay search like Super-KamioKande, demand an improved knowledge of the exclusive reaction channels such as strangeness associated productions. Therefore, our RPWIA results may play a significant role in the analysis of data from those experiments, and also in the separation process of backgrounds from the proton decay signal.

\section*{Acknowledgement}

This work is financially supported by German Academic Exchange Service (DAAD) in association with Africa Institute of Mathematical Sciences (AIMS) under contract number A-10-02802.

\end{document}